\documentstyle[eqsecnum,aps]{revtex}
\begin{document} 
\draft

\newcommand{\be}{\begin{equation}} 
\newcommand{\ee}{\end{equation}} 
\newcommand{\bea}{\begin{eqnarray}} 
\newcommand{\eea}{\end{eqnarray}}

\title{\hbox{ \rightline{ \normalsize  UFRJ--IF--FPC--001/98}} \Large 
\bf{Circulation Statistics in Three-Dimensional Turbulent Flows}} 
 
\author{ L. Moriconi$^{a)}$ and F.I. Takakura$^{b)}$} 
\address{$a)$ Instituto de F\'\i sica, Universidade Federal do Rio de Janeiro \\C.P. 68528, Rio de Janeiro, RJ --- 21945-970, Brasil} 
\address{$b)$ Departamento de F\'\i sica, ICE, Universidade Federal de Juiz de Fora,  
\\36036-330, Juiz de Fora, MG, Brasil} 
\maketitle 
 
\begin{abstract} 
We study the large $\lambda$ limit of the loop-dependent characteristic functional $Z( \lambda )=<\exp(i\lambda \oint_c \vec v \cdot d \vec x )>$, related to the probability density function (PDF) of the circulation around a closed contour $c$. The analysis is carried out in the framework of the Martin-Siggia-Rose field theory formulation of the turbulence problem, by means of the saddle-point technique. Axisymmetric instantons, labelled by the component $\sigma_{zz}$ of the strain field -- a partially annealed variable in our formalism -- are obtained for a circular loop in the $xy$ plane, with radius defined in the inertial range. Fluctuations of the velocity field around the saddle-point solutions are relevant, leading to the lorentzian asymptotic behavior $Z(\lambda) \sim 1/{\lambda^2}$. The ${\cal O}(1 / {\lambda^4})$ subleading correction and the asymmetry between right and left PDF tails due to parity breaking mechanisms are also investigated.   
\end{abstract}


\vskip 0.5 in
 
\section{Introduction} 
The study of the statistical properties of circulation in fully developed 
turbulence has been attracting a large deal of attention along the 
last few years \cite{migdal,cao,chorin}. The main motivation relies on the emergent picture of turbulence as a phenomenon intrinsically related to the dynamics of vorticity filaments, clearly observed for the first time in direct numerical simulations of the Navier-Stokes equations \cite{she1}. Filamentary structures seem to have, in fact, a fundamental place in the hierarchy of eddy fluctuations, as advanced in a recent phenomenological work of She and L\'ev\^eque \cite{she2}, where multifractal exponents of velocity structure functions were predicted to very accurate precision. 
 
An earlier theoretical analysis of the problem of circulation statistics was attempted by Migdal \cite{migdal}, who proposed, using functional methods originally devised for the investigation of gauge theories, that in the inertial range the PDF of the circulation $\Gamma$, ${\rm P} (\Gamma)$, evaluated for a closed contour $c$, should depend uniquely on the scaling variable $\Gamma / A^{(2k-1)/2k}$, where $A$ is the minimal area enclosed by $c$ and $k$ is an unknown parameter. It was initially thought, in order to compute $k$, that the central limit theorem could be evoked to regard $\Gamma$ as a random gaussian variable obtained from the contributions of many independent vortices. Using, then, the definition of circulation, 
\be 
\Gamma = \oint_c \vec v \cdot d \vec x \ , \ \label{i1} 
\ee 
and the Kolmogorov scaling law, $<|\vec v( \vec x)-\vec v( \vec y)|> \sim |\vec x- \vec y|^{1/3}$, a simple guess would be $k=3/2$, leading to $< \Gamma^n > \sim A^{2n/3}$. We now know, however, from a numerical analysis by Cao et al. \cite{cao} that although there is some support to the minimal area conjecture and the general existence of a scaling variable, as defined above, the gaussian description of the circulation statistics in the inertial range and the ``Kolmogorov" exponent $k=3/2$ are both ruled out
(the numerical results indicate $k < 3/2$). Gaussianity holds only in the integral scales, while it turns out that for loops contained in the inertial range the correlation between vortices cannot be neglected, a fact that obstructs an application of the law of large numbers. Intermittency, as found in ref. [2], is signaled at the tails of the circulation PDFs, which are fitted by stretched exponentials like ${\rm P}(\Gamma) \sim \exp \left( - \beta | \Gamma |^\alpha \right )$, where $\alpha \simeq 1$ in the inertial range and 
$\alpha \simeq 2$ in the integral scales. On the other hand, the circulation PDF cores are gaussian, as one could expect. 
 
An important conceptual point, raised in the same numerical simulation and related to the determination of $k$, is whether the moments of $\Gamma$ are independent or not from the form of velocity correlation functions. In order to find $< \Gamma^2 >$, for instance, it may be useless to know the two point correlation function $< v_\alpha (\vec x,t) v_\beta (\vec x',t)>$, since the contour integrations which appear in the definition of the square of the circulation and the average over realizations of the random velocity field may not commute. 
 
Our aim in the present work is to study the problem of circulation statistics in the inertial range through the Martin-Siggia-Rose (MSR) technique \cite{msr}. In spite of the many years passed since its advent, only recently interesting results were obtained from the MSR formalism, concerning the computation of intermittency effects in problems like turbulence in the Burgers model and in the transport of a passive scalar \cite{vik,falko}. The basic tool employed in these works is the saddle-point method, where instanton configurations and fluctuations around them are assumed to contribute in a significant way to the evaluation of the MSR functional. As we will see, a computation carried along these lines will give us non-gaussian tails for the circulation PDF, with stretching exponent $\alpha =1$, in reasonable agreement with the numerical findings commented above. 
 
This paper is organized as follows. In section II, the basic elements of our formalism are set. We define the MSR path integral expression from which the circulation PDF may be derived, and work out its instanton solutions. The saddle-point action is then computed. In section III, we move to the next natural step, which is the study of fluctuations around the saddle-point solutions. We find that fluctuations contribute in an essential way to the asymptotic form of the MSR functional. In section IV, we investigate subleading corrections to the asymptotic expression, induced by small (and gaussian) fluctuations of the circulation. As a result, we establish, for the PDF of the circulation, a relation between the width of its gaussian core and the tail decaying parameter $\beta$. In section V, we study the structure of asymmetric PDFs, due to parity breaking mechanisms, like turbulence in rotating systems or under the action of parity breaking external forces. We comment on our results in section VI, pointing out directions of future research. In the appendixes, we discuss in more detail computations which underly some of the results presented in the bulk of the paper.   
 
\section{Instantons in the MSR approach} 
As largely known, inertial range properties of three-dimensional turbulence may be modeled by the stochastic Navier-Stokes equations \cite{frisch},  
\bea 
& \partial_t & v_\alpha + v_\beta \partial_\beta v_\alpha = 
-\partial_\alpha P + \nu \partial^2 v_\alpha +f_\alpha \nonumber \ , \ \\ 
& \partial_\alpha & v_\alpha = 0 \ , \ \label{msr1} 
\eea 
where the $\alpha =1,2,3$ and the gaussian random force 
$f_\alpha (\vec x, t)$ is defined by 
\bea 
&<& f_\alpha (\vec x, t)> =0 \ , \ \nonumber \\ 
&<& f_\alpha (\vec x, t) f_\beta (\vec x',t') > \equiv 
D_{\alpha \beta}(\vec x - \vec x') \delta (t -t')=  
D_0 \exp \left ( -{|\vec x - \vec x'|^2 
\over L^2} \right ) \delta_{\alpha \beta} \delta (t - t') \ . \ \label{msr2} 
\eea 
Above, $L$ is the typical correlation length of the energy pumping mechanisms, acting at large scales. The other important length in the problem, according to Kolmogorov theory \cite{k41}, is $\eta \sim \nu^{3/4} \rightarrow 0$, the microscopic scale where viscosity effects come into play. 
 
From the stochastic Navier-Stokes equations one may try to obtain, in principle, any velocity correlation function. We are particularly interested to study the characteristic functional 
\be 
Z(\lambda)=<\exp \left( i \lambda \Gamma \right)> \ , \  \label{msr3} 
\ee 
where $\Gamma$ is the circulation evaluated at time t=0, as given by 
(\ref{i1}). The contour $c$ used in the definition of $\Gamma$ is taken here to be the circumference $x^2+y^2=R^2$, with $z=0$, oriented in the counterclockwise direction. A basic condition in our analysis is that $R$ is a length contained in the inertial range, that is, $\eta \ll R \ll L$. The PDF for the circulation may be written from the loop functional as 
\be 
{\rm P}(\Gamma)={1 \over {2 \pi}} \int^\infty_{- \infty} 
d \lambda \exp(-i \lambda \Gamma)Z(\lambda) \label{msr4} 
\ . \ 
\ee 
It is appropriate, for the computations which will follow, to consider the analytical mapping $\lambda \rightarrow -i \lambda$ in the RHS of (\ref{msr3}). At a later stage we will get back to the original definition of $\lambda$.  
The MSR formalism \cite{msr} allows us to write the path-integral expression 
\be 
Z(\lambda)= \int D \hat v Dv DP DQ \exp(-S) \ , \ \label{msr5} 
\ee 
where 
\bea 
S=&-&i \int d^3 \vec x dt \left [ \hat v_\alpha (\partial_t v_\alpha + v_\beta  \partial_\beta  v_\alpha -\nu \partial^2 v_\alpha 
+ \partial_\alpha P)+ Q \partial_\alpha v_\alpha \right ] 
\nonumber \\ 
&+&{1 \over 2} \int dt d^3 \vec x d^3 \vec x' \hat v_\alpha(\vec x,t) 
D_{\alpha \beta}(\vec x - \vec x') \hat v_\beta (\vec x', t) 
- \lambda \Gamma \ . \ \label{msr6} 
\eea

The MSR technique may be used to derive, in an alternative way, the Wyld
diagrammatic expansion \cite{wyld} for the computation of correlation
functions of the velocity field, obtained directly from the stochastic
equations (\ref{msr1}). This expansion is constructed by taking
the non-linear term in the Navier-Stokes equations, related to convection, as a perturbation.
For this reason, the perturbative MSR-Wyld approach has been criticized
along the years, as an inappropriate tool to deal with the singular
configurations of the velocity field, which are fundamental in turbulence.
However, the advantage of the MSR formalism is that non-perturbative issues
may be addressed in principle, if one knows how to find specific configurations
of the flow that could represent relevant contributions to the
functional integration for $Z(\lambda)$. This is precisely the task to which
the saddle-point method is devised for.

The role of the $P$ and $Q$ fields in the above path-integral summation is just to assure that
$ \partial_\alpha v_\alpha = \partial_\alpha \hat v_\alpha = 0$. 
These incompressibility conditions are in fact two of the four saddle-point equations obtained from the action $S$, viz., 
\bea 
{{\delta S} \over {\delta Q}} &=& \partial_\alpha v_\alpha = 0 \ , \ 
\label{msr7} \\ 
{{\delta S} \over {\delta P}} &=& \partial_\alpha \hat v_\alpha = 0 \ . \ \label{msr8} 
\eea 
The other two saddle-point equations are given by 
\bea 
&&{{\delta S} \over {\delta v_\alpha}} = 
i \left (\partial_t \hat v_\alpha - \hat v_\beta \partial_\alpha 
v_\beta + v_\beta \partial_\beta \hat v_\alpha + \nu \partial^2 \hat 
v_\alpha + \partial_\alpha Q \right) - \lambda  
{{\delta \Gamma} \over {\delta v_\alpha}} 
= 0 \ , \ \label{msr9} \\ 
&&{{\delta S} \over {\delta \hat v_\alpha}} = 
i \left ( \partial_t  v_\alpha + v_\beta \partial_\beta 
v_\alpha - \nu \partial^2 v_\alpha + \partial_\alpha P \right ) 
- \int d^3 \vec x' D_{\alpha \beta}(|\vec x -\vec x'|) 
\hat v_\beta (\vec x',t) 
= 0 \ . \ \label{msr10} 
\eea 
We have 
\be 
{{\delta \Gamma} \over {\delta v_\alpha}} = 
{{\delta} \over {\delta v_\alpha}} \oint_c v_\beta(\vec x',0) dx_\beta' =\epsilon_{3 \beta \alpha} {x_\beta \over r_\perp} \delta (r_\perp-R) \delta(z) \delta(t) \ , \ \label{msr11} 
\ee 
where $r_\perp =(x^2+y^2)^{1/2}$. The importance of the saddle-point equations is that they provide a systematic way to study the large $\lambda$ limit of $Z(\lambda)$. However, the saddle-point action computed in this way necessarily depends on $\lambda$ in a way incompatible with observational results \cite{cao}. In order to understand it, we observe that the saddle-point equations are invariant under the scaling transformations 
$ \nu \rightarrow h^{1/2} \nu $,  
$ t \rightarrow h^{-1/2} t $, 
$ v_\alpha \rightarrow h^{1/2} v_\alpha $, 
$ \hat v_\alpha \rightarrow h \hat  v_\alpha $, 
$ P \rightarrow h P $, 
$ Q \rightarrow h^{3/2} Q $ and 
$\lambda \rightarrow h \lambda$. 
These relations imply that the saddle-point action has the general form $S^{(0)} = \lambda^{3/2} f(\lambda^{-1/2} \nu)$. Since we expect to have 
finite answers in the limit of vanishing viscosity, it follows that 
$S^{(0)} \sim \lambda^{3/2}$. This dependence on $\lambda$ is exactly the 
one found in Burgers turbulence for the statistics of velocity differences \cite{vik,poly}, which we know not to reproduce, even qualitatively, the PDFs of the circulation in three-dimensions. A similar difficulty was in fact noticed in the investigation of velocity structure functions in incompressible turbulence by means of the saddle-point method \cite{falko}. In order to find physically meaningful PDF tails of the circulation, a solution of this problem will be pursued here, based on the definition of an additional field in the MSR path-integral, parametrizing an infinite family of saddle-point configurations.     
 
We would be tempted to study the above saddle-point equations by first eliminating the $P$ and $Q$ fields in (\ref{msr9}) and (\ref{msr10}) with the help of (\ref{msr7}) and (\ref{msr8}). All nonlinear terms in (\ref{msr9}) and (\ref{msr10}) would consequently appear projected on transverse modes through the use of the tensor 
$\Pi_{\alpha \beta} = \partial^{-2} (\partial_\alpha \partial_\beta -\delta_{\alpha \beta})$. However, this is not an adequate procedure to follow, in view of the simplifications inherent in the implementation of the saddle-point method to the MSR formalism. The central point is that we will be dealing with linear approximations for the velocity field, as a consequence of the small radius $R$ of the contour $c$, in comparison with the large scale length $L$. We have, thus, 
\be 
v_\alpha (\vec x ,t) = \sigma_{\alpha \beta}(t) x_\beta \ , \label{msr12} 
\ee 
with $\sum_\alpha \sigma_{\alpha \alpha}=0$ (due to $\partial_\alpha v_\alpha =0$). Coordinate independent terms are not written above, since we may impose, from invariance under the group of time-dependent translations, the saddle-point solution to satisfy $v_\alpha (\vec x =0,t)=0$ (see appendix A). Using (\ref{msr12}) we observe that expressions like $\Pi_{\alpha \beta} v_\gamma \partial_\gamma v_\beta$, related to the global nature of the flow, would not be precisely defined. A simple way out of this problem, usual in applied mathematical studies of the Navier-Stokes equations \cite{impa}, is to write the pressure as a quadratic form, 
\be 
P= A_{\alpha \beta} x_\alpha x_\beta \ , \  \label{msr13} 
\ee 
so that $ \partial_\alpha P $ exactly cancels in (\ref{msr10}) any symmetric tensor acting on the spatial coordinates, which would appear in the linear approximation. Therefore, (\ref{msr10}) may be written as an equation for the time evolution of the antisymmetric part of the strain field, 
\be 
{d \over { dt}} \sigma^{\bar s}_{\alpha \beta} + 
(\sigma^s \sigma^{\bar s} + \sigma^{\bar s} \sigma^s)_{\alpha \beta} - 
i \int d^3 \vec x \partial_{[\beta ,} D_{\alpha ] \gamma} 
(|\vec x|) \hat v_\gamma (\vec x ,t) = 0 \ , \  \label{msr14} 
\ee 
where we have defined 
\bea 
&& \sigma^s_{\alpha \beta} = {1 \over 2} ( \sigma_{\alpha \beta} 
+ \sigma_{\beta \alpha}) \ , \ \label{msr15} \\ 
&& \sigma^{\bar s}_{\alpha \beta} = {1 \over 2} ( \sigma_{\alpha \beta} 
- \sigma_{\beta \alpha}) \ , \ \label{msr16} \\ 
&& \partial_{[\beta ,} D_{\alpha ] \gamma} (|\vec x|)= {1 \over 2} 
\left ( \partial_\beta D_{\alpha  \gamma} (|\vec x|)- 
\partial_\alpha D_{\beta \gamma} (|\vec x|) \right ) \ . \ \label{msr17} 
\eea

An important remark is that (\ref{msr12}) is not assumed to represent a direct modeling of the velocity 
field in sustained turbulence, which we know to be associated with many different length scales and 
singular structures. The idea of the instanton method, as advanced by Falkovich et al. \cite{falko}, is in 
fact to consider, in the MSR framework, smooth configurations and perturbations around them that may 
condense some information on the statistics of the strong (intermittent) fluctuations of the velocity field. 
The situation here is analogous to the well-known instanton approach to the double well potential in 
quantum mechanics \cite{coleman}, where instantons are obtained as saddle-point solutions, yielding 
extremes of the euclidean action. It is clear in that case that the smooth kink/anti-kink form of the 
instanton configurations cannot be taken as a direct representation of the quantum-mechanical dynamics, 
which has a picture as a sum over particle paths with complex weights $\exp(iS)$. In the turbulence 
context, instead of transforming time into an imaginary variable as it is done quantum mechanics, we look 
for saddle-point solutions, considering, in the MSR action, the analytical mapping $\lambda \rightarrow -i 
\lambda$. A deeper analogy, which should also be noted, is provided by the phenomenon of localization in 
condensed matter physics. There is, in this case, a functional integral formalism, where smooth instantons 
may be found, giving expressions for the tails of the density of  electron states \cite{qhe}. The similarity 
with the turbulence problem is a strong one: while in the condensed matter system localized 
wavefunctions define some multifractal set, the same phenomenon takes place in turbulence, regarding 
the fluctuations of the velocity field. Also, the limitations of the instanton  method are exactly the same in 
both problems. Either in localization or in turbulence the core of the density  of states or of the PDFs, 
respectively, cannot be obtained from the saddle-point technique. To understand it in our analysis of the
statistics of circulation, we note that for large values of $\lambda$ the functional  $Z(\lambda)$
gets its more relevant contributions from the tails of the circulation PDF. At the core, where the PDF is essentially stationary, fluctuations of $\exp( i\lambda \Gamma)$ will tend to produce destructive interference. 
 
Our problem has been reduced so far to an analysis of equations (\ref{msr8}), (\ref{msr9}) and (\ref{msr14}), where in the second equation the velocity field is given by (\ref{msr12}). Since these equations are invariant under rotations around the $z$ axis, it is interesting to look for axisymmetric solutions. In the linear approximation, the most general form of an axisymmetric strain field is given by
\be 
\sigma (t) =
\left[
\matrix{ a(t) & b(t) & 0\cr
         -b(t) & a(t) & 0\cr
          0 & 0 & -2a(t)} \right] \ . \ \qquad
\label{msr18} 
\ee
The above form of $\sigma(t)$ has a simple hydrodynamical interpretation. Taking $a>0$, for instance, streamlines are
just expanding spirals which approach in an exponential way the $xy$ plane from both regions $z>0$ and $z<0$.
It is important to note that $\sigma_{z z} (t) = -2 a(t)$, which has the dimensions of the inverse of time, plays the role of an arbitrary external function in eq. (\ref{msr14}). In other words, vorticity is controlled by
stretch, associated to $a(t)$. We should try to find instantons (the solutions of the saddle-point equations) for 
any well-behaved function $a(t)$ (with $a(t) \rightarrow 0$ as $|t| \rightarrow \infty$) and then sum up their 
contributions in the path-integral expression for $Z(\lambda)$. This suggests an alternative strategy of 
computation, where $a(t)$, or some variable related to it, would appear from the very start in the MSR 
formalism as a field labelling families of velocity configurations. There are, in fact, many different ways to 
implement this idea, distinguished essentially by computational convenience. Our choice consists in writing 
(\ref{msr5}), up to a normalization factor, as 
\bea 
&&  Z(\lambda) = \int D \hat v Dv DP DQ D \sigma^s \delta 
\left[ \left. \partial_\alpha v_\beta \right |_{z=0} 
+ \left. \partial_\beta v_\alpha \right |_{z=0} - 2 \sigma^s_{\alpha \beta} \right] \exp(-S) \nonumber \\ 
&& = \int D \sigma^s \int D \hat v Dv DP DQ D \tilde Q \exp(- \tilde S) \ , \ \label{msr19} 
\eea 
where $ \sigma^s_{\alpha \beta}= \sigma^s_{\alpha \beta}(x,y,t)$ and
$\tilde Q_{\alpha \beta} = \tilde Q_{\alpha \beta}(x,y,t)$ are symmetric matrices and 
\be 
\tilde S = S - {i \over 2} \int dx dy dt \tilde Q_{\alpha \beta}(x,y,t) \cdot 
\left ( \left. \partial_\alpha  v_\beta  \right |_{z=0} 
+\left. \partial_\beta v_\alpha \right |_{z=0} - 2 \sigma^s_{\alpha \beta}(x,y,t) \right) \ . \ \label{msr20} 
\ee 
The meaning of (\ref{msr19}) is that we sum up the contributions to the path-integral expression in two steps: first by considering velocity configurations which satisfy $ \left. \partial_\alpha v_\beta \right |_{z=0}
+ \left. \partial_\beta v_\alpha \right |_{z=0} = 2 \sigma^s_{\alpha \beta}(x,y,t)$, for a given field
$\sigma^s_{\alpha \beta}$. The summation over the fields $\sigma^s_{\alpha \beta}$ is performed 
afterwards. The linear approximation for the velocity field corresponds, thus, to fields
 $\sigma^s_{\alpha \beta}$ with slow dependence on the $x$ and $y$ coordinates, within the length scale of 
the order of $R$, while axial symmetry, a condition related to large values of $\lambda$, is imposed here as 
a restriction on the configurations for $\sigma^s_{\alpha \beta}(t)$. More precisely, we will consider the sum 
in (\ref{msr19}) as carried over the space of axisymmetric fields 
$\sigma^s_{\alpha \beta}(t) = ( \delta_{\alpha \beta} - 3 \delta_{\alpha 3} \delta_{\beta 3})a(t)$,
in accordance with (\ref{msr18}). This corresponds to replacing $\int D \sigma^s(t)$ by $\int D a(t)$ in 
(\ref{msr19}). However, this constraint has to be applied with care, since its meaning is linked to 
configurations of the velocity field defined at length scales larger than the loop's radius $R$. To state it in a
different way, the velocity field that enters in the above delta functional is in fact a ``smeared" field, given by 
the contributions of wavenumbers $k< R^{-1}$.

The saddle-point method is to be used in the first step of computation (where 
$\sigma^s_{\alpha \beta}$ is fixed) involving the action $\tilde S$ rather than $S$.
The only modification of the previous saddle-point equations (\ref{msr7} - \ref{msr10}), as may be readily 
seen from $\tilde S$, is on (\ref{msr9}), which must be replaced now by 
\be 
{{\delta \tilde S} \over {\delta v_\alpha}} = 
i \left (\partial_t \hat v_\alpha - \hat v_\beta \partial_\alpha 
v_\beta + v_\beta \partial_\beta \hat v_\alpha + \nu \partial^2 \hat 
v_\alpha + \partial_\alpha Q +
\partial_\beta  ( \delta (z) \tilde Q_{\beta \alpha} ) \right ) - \lambda  
{{\delta \Gamma} \over {\delta v_\alpha}} 
= 0 \ . \ \label{msr21} 
\ee 
We also have an additional equation, associated to variations of the 
field $\tilde Q_{\alpha \beta}$, 
\be 
{{\delta \tilde S} \over {\delta \tilde Q_{\alpha \beta}}}=-i 
\left ( \left. \partial_\alpha v_\beta \right |_{z=0}
+\left. \partial_\beta v_\alpha \right |_{z=0} - 2 \sigma^s_{\alpha \beta}(t) \right)= 
0 \ . \  \label{msr22} 
\ee 
This equation, however, is beforehand solved by (\ref{msr12}) and (\ref{msr18}). Using (\ref{msr11}), (\ref{msr12}) and taking the limit of vanishing viscosity, we may write (\ref{msr21}) as 
\bea 
&& \partial_t \hat v_\alpha - \sigma_{\beta \alpha} \hat v_\beta + \sigma_{\beta \gamma} x_{\gamma} 
\partial_\beta \hat v_\alpha + \partial_\alpha Q +  \partial_\beta
(\delta (z)  \tilde Q_{\beta \alpha}) = \nonumber \\ 
&& = i\lambda \epsilon_{3 \alpha \beta} {x_\beta \over r_\perp} \delta (r_\perp-R) \delta(z) \delta(t) \ . \ 
\label{msr23}  
\eea 
We have, therefore, a closed system of coupled equations, given by 
(\ref{msr8}),(\ref{msr14}) and (\ref{msr23}). It is important to state the boundary conditions that the 
solutions of these equations have to satisfy. Since the viscosity term appears in (\ref{msr21}) with the 
opposite sign, compared to the one in the Navier-Stokes equations, we impose, in order to avoid an 
unbounded growing of the field $\hat v_\alpha (\vec x,t)$, that 
$\hat v_\alpha(\vec x, t>0)=0$. 
In this way, (\ref{msr23}) leads us to 
\be 
\hat v_\alpha(\vec x, 0^-)=  
i\lambda \epsilon_{3 \beta \alpha} {x_\beta \over r_\perp} \delta (r_\perp-R) \delta(z) \ . \ \label{msr24} 
\ee 
Also, we require that $\hat  v_\alpha (\vec x , t) \rightarrow 0$ as $t \rightarrow - \infty$. The equation for $\hat 
v_\alpha (\vec x,t)$ may be solved through the ansatz 
\be 
\hat v_\alpha (\vec x, t) = 
\epsilon_{3 \beta \alpha} x_\beta \delta (z) 
\sum_{n=0}^{\infty} c_n(t) r_\perp^{n-1}\delta^{(n)} (r_\perp-R) \ , \ 
\label{msr25} 
\ee 
where $\delta^{(n)}(r_\perp-R) = d^n \delta(r_\perp-R) / dr_\perp^n$. The boundary condition (\ref{msr24}) 
reads now 
\bea 
&&c_0(0^-)=i \lambda \ , \ \nonumber \\ 
&&c_n(0^-)=0,{\hbox{ for $n > 0$}} \ . \ \label{msr26} 
\eea 
We find, substituting (\ref{msr25}) in (\ref{msr23}), 
\bea 
&& {d \over {dt}} c_0 +ac_0= 0 \ , \ \nonumber \\ 
&& {d \over {dt}} c_n +a(n+1)c_n+ac_{n-1} = 0,{\hbox{ for $n > 0$}} 
\label{msr27} 
\eea 
and $\tilde Q_{\alpha \beta} = (\delta_{\alpha \beta} - \delta_{\alpha 3} \delta_{\beta 3}) \tilde Q$, with
(below, $\alpha =1,2$) 
\bea 
&& \partial_\alpha \tilde Q = - 2b(t)x_\alpha  
\sum_{n=0}^{\infty} c_n(t) r_\perp^{n-1}\delta^{(n)} (r_\perp-R) \ , \ \label{msr28} \\ 
&&Q=0 \ . \ \label{msr29} 
\eea 
The infinite set of equations (\ref{msr27}) as well as (\ref{msr28}) are solved respectively by  
\bea 
&&c_n(t)={{i \lambda} \over {n!}} e^{ -\int_0^t dt' a(t')} 
\left( e^{ -\int_0^t dt' a(t')} -1 \right)^n \ , \ \label{msr30} \\ 
&& \tilde Q (r_\perp,t) = - 2b(t) \sum_{n=0}^{\infty} c_n(t) 
\int_0^{ r_\perp} d \xi \xi^n \delta^{(n)} ( \xi- R ) 
=  - 2 i \lambda b(t) \theta (r_\perp - R e^{ \int_0^t dt' a(t')}) \ , \ \label{msr31} 
\eea 
where $\theta (x) \equiv (1+ |x|/x)/2$ is the step function. Taking (\ref{msr30}), the infinite summation in 
(\ref{msr25}) may be exactly performed. We find the compact result for $t<0$, 
\be 
\hat v_\alpha(\vec x,t)=  
i\lambda \epsilon_{3 \beta \alpha} { x_\beta \over r_\perp} \delta (r_\perp-R  e^{ \int_0^t dt' a(t')}) \delta(z) 
\ . \ \label{msr32} 
\ee
In order to get some intuition on the singularity in the above expression, we just recall that the quadratic term 
for
$\hat v_\alpha (\vec x,t)$ in the MSR action is obtained from
\be
< \exp(i \int d^3 \vec x dt \hat v_\alpha (\vec x,t) f_\alpha (\vec x,t))>_f \ , \ \label{o1}
\ee
where the brackets denote an average over realizations of the stochastic force field $f_\alpha (\vec x,t)$.
Substituting in this average $\hat v_\alpha (\vec x,t)$ by the saddle-point solution (\ref{msr32}), we find
\be
\int d^3 \vec x dt \hat v_\alpha (\vec x,t) f_\alpha (\vec x,t) \sim \int dt \oint dx_\alpha f_\alpha (\vec x,t) \ , \
\label{o2}
\ee
where the loop integral is taken around the circumference of radius $r_\perp = R \exp (\int_0^t dt' a(t'))$.
We see that (\ref{o2}) is in fact non-vanishing for configurations of the force field that may produce some circulation
around the loop  $r_\perp = R$, at $t=0$, through convective processes in the fluid.

Let us consider now equation (\ref{msr14}) for the velocity field, which, using the strain field (\ref{msr18}), may be written as 
\be 
\dot b + 2ab +i \int d^3 
\vec x \partial_{[ 1 ,} D_{ 2 ] \gamma} 
(|\vec x|) \hat v_\gamma (\vec x ,t) = 0 \ . \ \label{msr33} 
\ee 
Substituting the solution for $\hat v_\gamma (\vec x, t)$, in the above expression, we obtain 
\be 
\dot b + 2ab = -2 \pi  D_0 \lambda \left ( {R \over L} \right )^2 
 e^{2 \int_0^t dt' a(t')} \theta (-t) \ . \ \label{msr34} 
\ee 
In order to have well-behaved solutions for $t \rightarrow - \infty$, 
we see, from (\ref{msr34}), that it is necessary to have 
in this limit $\int_0^t dt' a(t') \rightarrow - \infty $. 
Motivated by the general idea of a gradient expansion, we will restrict our study, as a first approximation, to the effects of time independent configurations given by $a(t)=a>0$. Correspondingly, in the definition of $Z( \lambda )$, eq. (\ref{msr19}), we will have 
\be 
\int D \sigma^s \rightarrow \int_0^{ \infty} da \ . \ \label{msr35} 
\ee 
A possible physical interpretation of the above replacement is related to the experimental observation of circulation as a more intermittent random variable than longitudinal velocity differences \cite{cao}. Thus, in the decomposition of the strain tensor (\ref{msr18}) into symmetric and antisymmetric parts, the latter is actually the quantity which fluctuates more strongly in the ``background" defined by the partially annealed field $a(t)$. It is worth observing this kind of interpretation is usual in a large variety of systems characterized by different time scales, like spin glasses, for instance, in the situation where the dynamics of spin couplings is slow -- but not negligible -- when compared to the typical time for spins to reach thermal equilibrium \cite{dot}. 
 
Equation (\ref{msr34}) may be easily solved, yielding 
\be  
b(t)=-{{ \pi D_0 \lambda} \over {2a}} 
\left ( {R \over L} \right )^2 e^{-2a |t|} \ . \ \label{msr36} \\ 
\ee
As it could be anticipated, we see that (\ref{msr36}) represents the well-known phenomenon of
vorticity amplification by vortex stretching, controlled by the parameter $a$. Although viscosity does not 
enter in this expression, vortex stretching is bounded, which would not occur in an inviscid flow. The 
explanation for this behavior of the instanton solution follows from the fact that viscosity has been taken 
into account in an implicit way, through eq. (\ref{msr24}), which defines $\hat v_\alpha (\vec x,t)$ at the 
initial time $t=0$, so that the saddle-point solutions vanish as $t \rightarrow \pm \infty$. The peculiar property of (\ref{msr36}) that will be important in our subsequent considerations is just the factor $\lambda / a$, relating  $\lambda$ and the vortex stretching 
parameter $a$ to the amplitude of $b(t)$. 

The saddle-point solutions we have found for $v_\alpha (\vec x,t)$ and $\hat v_\alpha (\vec x,t)$  may be substituted now in the action $ \tilde S$ to give 
\be 
\tilde S^{(0)} = -{{ \pi^2 D_0 R^4} \over {2 L^2}} \cdot {{\lambda^2} \over a} 
\ . \ \label{msr37} 
\ee 
We note that a straight application of this result would lead to 
\be 
Z( \lambda) \sim \int_0^{\infty} da  
\exp( -{{ \pi^2 D_0 R^4} \over {2 L^2}} \cdot {{\lambda^2} \over a} ) 
\ , \ \label{msr38} 
\ee 
which is divergent as the integration region extends to 
$a \rightarrow  \infty$ (above, $\lambda$ has been substituted by 
$i \lambda$). This ``ultraviolet" divergence is in fact naturally regularized when we also take into account fluctuations around the saddle-point solutions, as shown next. 
 
\section{Analysis of fluctuations}
Denoting the saddle-point fields and fluctuations around them by the indexes $``(0)"$ and $``(1)"$, respectively, we write 
\bea 
&& v_\alpha (\vec x, t) = v_\alpha^{(0)} (\vec x, t)+ 
v_\alpha^{(1)} (\vec x, t) \ , \ 
\hat v_\alpha (\vec x, t) = \hat v_\alpha^{(0)} (\vec x, t)+ 
\hat v_\alpha^{(1)} (\vec x, t) \ , \ \nonumber \\ 
&&P(\vec x, t) = P^{(0)}(\vec x, t) + P^{(1)}(\vec x, t) \ , \  Q(\vec x, t) = Q^{(0)}(\vec x, t) + Q^{(1)}(\vec x, t) \ , \ \nonumber \\ 
&&\tilde Q_{\alpha \beta}(x,y,t) = \tilde Q^{(0)}_{\alpha \beta}(x,y,t) + \tilde Q^{(1)}_{\alpha \beta}(x,y,t) \ . \ \label{af1} 
\eea 
The action is expressed as $\tilde S = \tilde S^{(0)}+ \tilde S^{(1)}$, where $\tilde S^{(0)}$ is given by (\ref{msr37}), and we have, up to second order in the perturbations, 
\bea 
\tilde S^{(1)} = &-& i \int d^3 \vec x dt \left [ \hat v^{(1)}_\alpha (\partial_t v^{(1)}_\alpha + v^{(0)}_\beta  \partial_\beta  v^{(1)}_\alpha + v^{(1)}_\beta  \partial_\beta  v^{(0)}_\alpha  - \nu \partial^2 v^{(1)}_\alpha + \partial_\alpha P^{(1)}) \right. \nonumber \\ 
&+& \hat v_\alpha^{(0)} 
( v^{(1)}_\beta \partial_\beta v^{(1)}_\alpha ) 
+ Q^{(1)} \partial_\alpha \left. v^{(1)}_\alpha \right ] 
- {i \over 2} \int d  x d y dt \tilde Q^{(1)}_{\alpha \beta} 
\left (  \partial_\alpha v_\beta^{(1)} +\left. \partial_\beta v_\alpha^{(1)} \right )
\right |_{z=0} \nonumber \\ 
&+& {1 \over 2} \int dt d^3 \vec x d^3 \vec x' \hat v^{(1)}_\alpha(\vec x,t) 
D_{\alpha \beta}(\vec x - \vec x') \hat v^{(1)}_\beta (\vec x', t) \ . \ 
\label{af2} 
\eea 
We included in (\ref{af2}), for the sake of completeness, the viscosity term, which in fact will be assumed to vanish in the next computations (nevertheless, we have to keep in mind that viscosity, as discussed before, plays an important role in the choice of the boundary condition for $\hat v^{(0)}_\alpha (\vec x,t)$ at $t=0$).

The integrations over $P^{(1)}$, $Q^{(1)}$ and $\tilde Q^{(1)}_{\alpha \beta}$ imply that 
\bea 
&& {\hbox{a)   }} \partial_\alpha v^{(1)}_\alpha (\vec x, t) =0 \ , \ \nonumber \\ 
&& {\hbox{b)   }} \partial_\alpha \hat v^{(1)}_\alpha (\vec x, t) =0 \ , \ \nonumber \\ 
&& {\hbox{c)   }} \left ( \left. \partial_\alpha v^{(1)}_\beta (\vec x, t) + 
\partial_\beta  v^{(1)}_\alpha (\vec x, t) \right ) \right |_{z=0} =0 \ . \ \label{af3} 
\eea 
If perturbations are written in a form which satisfy these relations, as we will do, then the fields $P^{(1)}$, $Q^{(1)}$ and $\tilde Q^{(1)}_{\alpha \beta}$ may be taken out from $\tilde S^{(1)}$. We are interested to find expressions for $v^{(1)}_\alpha (\vec x,t)$ and $\hat v^{(1)}_\alpha (\vec x,t)$, which describe effective degrees of freedom. 
 
The singularity of $\hat v_\alpha^{(0)} (\vec x,t)$ at $r_\perp = Re^{at}$ 
and $z=0$, given by (\ref{msr32}), represents a ring that shrinks to a point 
as $t \rightarrow - \infty$. One could imagine local fluctuations around 
$\hat v_\alpha^{(0)} (\vec x,t)$ given by variations of the vector field
defined on the ring,
\be
\hat v^{(1)}_\alpha (\vec x ,t) = \varphi_\alpha (\theta,t) \delta (r_\perp
-R e^{at}) \delta (z) \ , \  \label{af4}
\ee
where $ \theta$ is the azimuthal angle in cylindrical coordinates. Since $\varphi_\alpha (\theta,t)=\varphi_\alpha (\theta + 2 \pi,t)$, we may write
the Fourier series $\varphi_\alpha (\theta,t)=
\sum_{n= -\infty}^\infty \varphi_\alpha^{(n)} (t) \exp (i n \theta)$.
The incompressibility condition (\ref{af3}b), however, implies that $\varphi_\alpha^{(n)}=0$, for $n \not = 0$, and $ \varphi_\alpha^{(0)} (t) \equiv \delta c(t) \epsilon_{3 \alpha \beta} x_\beta / r_\perp $. Therefore, we are only allowed to consider amplitude fluctuations as
\be 
\hat v_\alpha^{(1)} (\vec x,t) = \delta c(t) \epsilon_{3 \alpha \beta} 
{x_\beta \over r_\perp} \delta(r_\perp -Re^{at}) \delta(z) \ . \ \label{af5} 
\ee 
An important remark is that the above expression is valid exclusively for negative times, since  $\hat v_\alpha^{(0)} (\vec x,t>0) = 0$.

We could also take into account perturbations of the ring that would deform its shape, but a little reflection
shows they may be neglected. Consider, for instance, perturbations of the ring in the $xy$ plane,
given by a field $\eta (\theta, t)$:
\be
\hat v_\alpha (\vec x,t)  = \varphi_\alpha (r_\perp,\theta; \eta )
 \delta(r_\perp -Re^{at}+ \eta (\theta,t)) \delta(z) \ ,\ \label{e1} 
\ee
where the above amplitude $\varphi_\alpha$ is a functional of $\eta( \theta,t)$ and satisfies to
$\varphi_\alpha  (r_\perp,\theta; \eta=0)=i \lambda \epsilon_{3 \alpha \beta} x_\beta / r_\perp$.
Up to first order in $\eta (\theta,t)$ we may write
\be
\hat v_\alpha^{(1)} (\vec x,t) = \left [
\int d \theta ' \eta (\theta ',t) {\delta \over {\delta \eta (\theta ',t)}}
\varphi_\alpha (r_\perp,\theta; \eta=0 ) \right ] 
\delta (r_\perp -Re^{at}) \delta(z) + i \lambda \epsilon_{3 \alpha \beta} 
{x_\beta \over r_\perp} \eta (\theta,t) \delta^{(1)}(r_\perp -Re^{at}) \delta(z) \ . \ \label{e2} 
\ee
The first term in the RHS of this equation may be absorbed by fluctuations given by (\ref{af4}). Regarding the second term, the same steps that led to (\ref{af5}) give us now  $\partial_\theta \eta (\theta,t)=0$, that is, the ring is deformed in the $xy$ plane through uniform radius variations. It is clear, due to the derivative of the delta function in (\ref{e2}) that (\ref{af5}) is in fact a more relevant contribution at lower wavenumbers. The same reasoning may be extended to generic perturbations of the ring's shape. The approximation of neglecting deformations of the ring would be inconsistent if there were small scale fluctuations of the velocity field taking place in a neighborhood of the ring, as we would conclude from the coupling of type $\hat v v$ in the action (\ref{af2}). However, as it will be shown in a moment, small scale fluctuations of the velocity field are contained only in some small compact region surrounding the origin. 

In view of the action of random forces at large length scales
($k < L^{-1}$, in Fourier space),
we keep, as a first approximation, the linear dependence of the velocity field on the spatial coordinates,
introducing fluctuations of the strain field as
\be 
v_\alpha^{(1)} (\vec x,t) =  a_\alpha (t)+ \omega_\beta (t) \epsilon_{\alpha \beta \gamma} x_\gamma \ . \ 
\label{af6} 
\ee
This linear expression is the only one compatible with the constraints (\ref{af3}a) and (\ref{af3}c). 

If we take $\delta c(t) =$ constant, it is not difficult to see, substituting (\ref{af5}) and (\ref{af6}) in
(\ref{af2}), that $\tilde S^{(1)}$ will not depend on $a_\alpha (t)$ or $\omega_\beta (t)$ for
$t < 0$. In other words, we have defined a ``zero mode" configuration, which would render the MSR
path-integral completely independent on large scale fluctuations of the velocity field. The solution of this 
problem consists in considering generic time-dependent variations $\delta c(t)$, precisely as we are doing, 
in accordance with the usual procedure for the treatment of zero modes associated to instantons 
\cite{coleman}.

Relations (\ref{af5}) and (\ref{af6}) were both defined through arguments based on the assumption that fluctuations around the saddle-point have to be local.
We observe, however, that they do not exhaust, in principle, the effective form of perturbations, which may occur also at smaller length scales.
In order to achieve full expressions for $\hat v_\alpha^{(1)} (\vec x,t)$ and
$v_\alpha^{(1)} (\vec x,t)$, it is necessary to take a closer look at
fluctuations associated to the dynamics of the action $\tilde S^{(1)}$.
Disregarding the coupling
$\hat v_\alpha^{(0)} ( v^{(1)}_\beta \partial_\beta v^{(1)}_\alpha )$ 
-- a self-consistent approximation, as we will see -- one may note that $\tilde S^{(1)}$, which governs the random behavior of $v^{(1)}_\alpha (\vec x,t)$, is the MSR field theory obtained from the stochastic equations 
\be 
\partial_t v^{(1)}_\alpha + v^{(0)}_\beta \partial_\beta v^{(1)}_\alpha + v^{(1)}_\beta \partial_\beta v^{(0)}_\alpha = 
\nu \partial^2 v^{(1)}_\alpha 
- \partial_\alpha P^{(1)} + f^{(1)}_\alpha \label{af7} 
\ee 
and the constraints (\ref{af3}a) and (\ref{af3}c). The random force
$f^{(1)}_\alpha (\vec x,t)$, like $f_\alpha (\vec x,t)$, is defined by
(\ref{msr2}).
A criterion to find the region of space where small scale fluctuations
determined by (\ref{af7}) may effectively occur is based on an analysis of the
local power supplied to the fluid by the pressure and external forces.
In the absence of perturbations, the laminar flow is described by the velocity
field $v^{(0)}_\alpha (\vec x,t)$, with power density
\bea 
{\cal P}_0 &=& v^{(0)}_\alpha (\vec x,t) \left ( - \partial_\alpha P^{(0)}(\vec x ,t)
+i x_\beta \int d^3 \vec x' \partial_{[\beta ,} D_{\alpha ] \gamma} 
(|\vec x'|) \hat v^{(0)}_\gamma (\vec x' ,t) \right ) \nonumber \\ 
&=& (a^2 + 3b(t)^2)ar^2_\perp - 8a^3z^2 \ , \ \label{af8} 
\eea 
where $b(t)$ is given by (\ref{msr36}) and $P^{(0)}$ is obtained according to the discussion which leads to eq. (\ref{msr14}). 
Taking $a > (D_0 \lambda)^{1/2}$, the $b(t)^2$ term in the above equation may be neglected. We get 
\be 
{\cal P}_0 \simeq a^3r^2_\perp - 8a^3z^2 \ . \ \label{af9} 
\ee 
The lower bound $(D_0 \lambda)^{1/2}$ for $a$ does not modify the asymptotic form of $Z(\lambda)$. We may check it by considering any regularized version of (\ref{msr38}), assuming its measure of integration is still dominated by the exponential factor as $a \rightarrow 0$. A more physical view on the lower bound for $a$, which will become clear later, is that in
order to evaluate the MSR functional $Z(\lambda)$, it is enough to take into account saddle-point configurations which
have support in the time interval $\Delta t \leq (D_0 \lambda)^{-1/2}$, so that the power density (\ref{af8})
turns out to be dominated by the symmetric part of the strain field.
 
The extra supply of power density provided by the pressure $P^{(1)}$ and the stochastic force $f^{(1)}_\alpha$ is 
\be 
{\cal P}_1 = < v^{(1)}_\alpha \left ( - \partial_\alpha P^{(1)} + 
f^{(1)}_\alpha \right ) > \ . \ \label{af10} 
\ee 
Since the equations and constraints for $ v_\alpha^{(1)} (\vec x,t)$ are linear, they are invariant under the
substitutions 
\be 
v_\alpha^{(1)} (\vec x,t) \rightarrow D^{1/2}_0 v_\alpha^{(1)} (\vec x,t) 
\ , \  
P^{(1)} (\vec x,t) \rightarrow D^{1/2}_0 P^{(1)} (\vec x,t) \ , \ 
f^{(1)}_\alpha (\vec x,t) \rightarrow D^{1/2}_0 f^{(1)}_\alpha (\vec x,t) \ . \ \label{af11} 
\ee 
The factor $D_0$ which appears in the two-point correlation function of the random force $f^{(1)}_\alpha$ is now replaced by unity. From eq. (\ref{af7}) and (\ref{af10}) we get, taking $\nu \rightarrow 0$,
\be
D_0^{-1} {\cal P}_1 = {1 \over 2} v_\beta^{(0)} \partial_\beta ( < v_\alpha^{(1)} v_\alpha^{(1)}>)
+{1 \over 2} (\partial_\beta v^{(0)}_\alpha + \partial_\alpha v^{(0)}_\beta)<v^{(1)}_\alpha v^{(1)}_\beta> \ . \
\label{h1}
\ee
At $\vec x =0$ we obtain
\be
D_0^{-1} {\cal P}_1 (\vec x =0) = (\delta_{\alpha \beta} - 3 \delta_{\alpha 3} \delta_{\beta 3}) a
<v^{(1)}_\alpha (0) v^{(1)}_\beta(0)> \ . \ \label{h2}
\ee
Since $a> (D_0 \lambda)^{1/2}$, we have $v^{(0)}_\alpha (\vec x,t) \simeq a(x_\alpha -3 \delta_{\alpha 
3}z)$, which means that the stochastic equation (\ref{af7}) involves essentially only two dimensional 
parameters: $a$ and $L$. Through simple dimensional analysis we may write
\be
<v^{(1)}_\alpha (\vec x) v^{(1)}_\beta (\vec x)> \equiv { C_{\alpha \beta} \over a} \ , \ \label{h3}
\ee
where $C_{\alpha \beta}$ is a dimensionless constant. We find, from (\ref{h2}) and (\ref{h3}),
\be
{\cal P}_1 =cD_0 \ , \ \label{h4}
\ee
where $c \equiv  C_{11} + C_{22} - 2 C_{33}$. From rotation symmetry around the $z$ axis, we have
$C_{11}=C_{22}$, and consequently $c= 2(C_{11}-C_{33})$. Due to the strong anisotropy in the system described by (\ref{af7}), we expect to have $c \neq 0$.

Considering now $|\vec x| \neq  0$, we may use dimensional analysis
once more to write for the first term in the RHS of (\ref{h1}),
\be
{1 \over 2} v_\beta^{(0)} \partial_\beta ( < v_\alpha^{(1)} v_\alpha^{(1)}>) \sim (x_\alpha - 3 \delta_{\alpha 3}z)
{C_\alpha \over L} \ , \ \label{h5}
\ee
where $C_\alpha$ is a dimensionless constant. Thus, for $|\vec x| \ll L$, the RHS of (\ref{h1}) is still 
dominated by the second term, leading us again to (\ref{h4}). It is important to observe that in the 
analysis presented above, the derivative in (\ref{h5}) is assumed to be a smooth function of the spatial 
coordinates, a condition that may be not valid in some specific set of points, as in a vortex sheet.

Eq. (\ref{h4}) is in fact a result similar to the one that would be obtained from a loose application of
Novikov's theorem \cite{novikov}.
We expect stronger fluctuations of the velocity
field for positions where $|{\cal P}_0| < |{\cal P}_1|$, that is \cite{comment}, 
\be 
|a^3r^2_\perp - 8a^3z^2| < |c| D_0 \ . \ \label{af12} 
\ee 
The above inequality is satisfied in a region of space bounded by three disjoint surfaces generated by the revolution of hiperbolae, as shown in fig.1. It is consistent to assume the surfaces have a well-defined meaning only at length scales contained in the inertial range. Since $a > (D_0 \lambda)^{1/2}$, we can see that for large enough values of $\lambda$, the surfaces enclose some region $\Omega$ surrounding the origin, with typical size $R_0 \sim (|c| D_0 / a^3)^{1/2} \ll R$. The condition on $\lambda$ is given by
\be 
{c^2 \over { \lambda^3 D_0 }} \ll R^4  
\ . \ \label{af13} 
\ee 
This relation defines, therefore, what is meant by the ``large
$ \lambda $ asymptotic limit".

To construct an effective picture out of these considerations, we imagine that in $\Omega$ additional fluctuations of $\hat v_\alpha^{(1)} (\vec x,t)$ and $v_\alpha^{(1)}(\vec x,t)$ are superimposed to the previous expressions (\ref{af5}) and (\ref{af6}). Physical results are then obtained in the $R_0/R \rightarrow 0$ limit. In practical terms, this amounts to rewriting $\tilde S^{(1)}$ in a form which explicitly takes into account the length scales involved here, $R_0$ and $R$. With this aim in mind, it is useful to employ the following notation: 
\be 
\hat v_\alpha^{(1)} (\vec x,t) =
\cases{ \hat v^{<}_\alpha (\vec x,t), & if $\vec x \not \in \Omega$\cr
         \hat v^{>}_\alpha (\vec x,t), & otherwise.\cr} 
\label{af13b} 
\ee 
Analogous definitions are provided for $v^{(1)}_\alpha (\vec x,t)$. 
We get, from (\ref{af13b}) and (\ref{af2}), 
\bea 
&\tilde S^{(1)}& = - i \int_{\vec x \in \Omega} d^3 \vec x dt  \hat v^{>}_\alpha (\partial_t v^{>}_\alpha + 
v^{(0)}_\beta \partial_\beta v^{>}_\alpha + v^{>}_\beta \partial_\beta v^{(0)}_\alpha ) 
- i \int_{\vec x \not \in \Omega} d^3 \vec x dt  \hat v^{<}_\alpha (\partial_t v^{<}_\alpha  \nonumber \\ 
&+& v^{(0)}_\beta \partial_\beta v^{<}_\alpha + v^{<}_\beta \partial_\beta v^{(0)}_\alpha ) 
+ {1 \over 2} \int_{\vec x,\vec x' \in \Omega} dt d^3 \vec x d^3  
\vec x' \hat v^{>}_\alpha(\vec x,t) 
D_{\alpha \beta}(\vec x - \vec x') \hat v^{>}_\beta (\vec x', t) 
\nonumber \\ 
&+& {1 \over 2} \int_{\vec x, \vec x' \not \in \Omega} dt d^3 \vec x   d^3 \vec x' \hat v^{<}_\alpha(\vec x,t) 
D_{\alpha \beta}(\vec x - \vec x') \hat v^{<}_\beta (\vec x', t) 
+ \int_{\vec x \in \Omega , \vec x' \not \in \Omega} dt d^3 \vec x 
d^3 \vec x' \hat v^{>}_\alpha(\vec x,t) 
\nonumber \\ 
& \times & D_{\alpha \beta}(\vec x - \vec x') \hat v^{<}_\beta 
(\vec x', t) 
\ . \ 
\label{af14} 
\eea 
According to the above discussion, we take now $\hat v_\alpha^{<} (\vec x,t)$ and $v_\alpha^{<} (\vec 
x,t)$ to be given by the former expressions (\ref{af5}) and (\ref{af6}), respectively. On the other hand, at 
smaller length scales, given by $|\vec x| < R_0$, (\ref{af6}) is not expected to reproduce
the behavior of $v_\alpha^{(1)} (\vec x,t)$ anymore, so that another
parametrization is needed, viz., 
\be 
v_\alpha^{>} (\vec x,t) =  \bar a_\alpha (t) + b_{\alpha \beta} (t) x_\beta 
\ . \ \label{af15} 
\ee
The linear expressions for (\ref{af6}) and (\ref{af15}) are associated to
the fact that we are considering velocity fluctuations to depend essentially 
on wavenumbers given by $k < L^{-1}$ and $k \sim R_0^{-1}$. 
Eq. (\ref{af15}) is not constrained by condition (\ref{af3}c), since it describes fluctuations at length scales 
$R_0 \ll R$. The surface $\partial \Omega$ which encloses $\Omega$ may be viewed as a vortex sheet for 
the velocity field $v^{(1)}_\alpha (\vec x,t)$. In appendix B, it is shown that $\Omega$ is necessarily a 
sphere of radius $R_0$, whereas $b_{\alpha \beta} (t)$ is an antisymmetric tensor and $a_\alpha (t) =
\bar a_\alpha (t)$. As the coordinate independent field $a_\alpha (t)$ ($= \bar a_\alpha (t)$) may be 
absorbed by pressure flucutations in the action (\ref{af2}), we may take
 \bea 
&&v_\alpha^{<} (\vec x,t) = \omega_\beta (t)\epsilon_{\alpha \beta \gamma} x_\gamma 
\ , \ \nonumber \\ 
&&v_\alpha^{>} (\vec x,t) = \phi_\beta (t) \epsilon_{\alpha \beta \gamma} x_\gamma 
\ , \ \label{af16} 
\eea 
where $\omega_\beta (t)$ and $ \phi_\beta (t)$ are proportional to the vorticity outside and inside $\Omega$, respectively. At this point we note that (\ref{msr32}) and (\ref{af16}) give 
\be 
\int d^3 \vec x \hat v_\alpha^{(0)} ( v^{(1)}_\beta \partial_\beta v^{(1)}_\alpha) =
\int_{\vec x \not \in \Omega} d^3 \vec x \hat v_\alpha^{(0)} ( v^{<}_\beta \partial_\beta v^{<}_\alpha )= 
0 \ , \ \label{af17} 
\ee 
proving the self-consistency of the simplification discussed before eq. (\ref{af7}).

From (\ref{af16}) we see that $v^{(1)}_\alpha (\vec x,t)$ gives no stretch.
This peculiar result is related to the fact that velocity fluctuations at length scales larger than $R$ have to 
satisfy both the constraints (\ref{af3}a) and (\ref{af3}c), which makes the flow described by (\ref{af7}) 
somewhat unusual, when  compared to the ones commonly modeled in fluid dynamics, where condition 
(\ref{af3}c) is not imposed. On a more physical ground, we may say the constraint (\ref{af3}c) means that 
the symmetric part of the strain  field is ``frozen'' and does not fluctuate around the saddle-point solution, 
which is a natural assumption, since we take it to represent the slow degrees of freedom. We also note that 
there is no contradiction between (\ref{h4}) and (\ref{af16}), since a coordinate independent field, as 
commented before, is not written explicitly for $v^<_\alpha (\vec x,t)$ and $v^>_\alpha (\vec x,t)$.

We found expressions for $v^{<}_\alpha (\vec x,t)$,
$v^{>}_\alpha (\vec x,t)$ and $\hat v^{<}_\alpha (\vec x,t)$, but nothing was said about $\hat v^{>}_\alpha (\vec x,t)$. As a matter of fact, this field will be replaced, as shown below, by linear combination of its moments $c_{\alpha \beta} (t) \equiv \int d \vec x \hat v^{>}_\alpha (\vec x,t) x_\beta$.

Substituting (\ref{af5}) ($= \hat v_\alpha^{<} (\vec x,t)$) and (\ref{af16}) in (\ref{af14}), we find, after a lengthy and straightforward computation,
\bea 
&&Z( \lambda ) \sim \int_{(D_0 \lambda)^{1/2}}^\infty 
da \int D[ \delta c(t)] D[ \rho (t)] \prod_{\alpha=1}^3 
D[ c_\alpha (t)] D[ \phi_\alpha (t)]  \nonumber \\ 
&& \exp  \{ -{{ \pi^2 D_0 R^4} \over {2 L^2}} \cdot {{\lambda^2} \over a} + 2i \int_{- \infty}^{\infty} dt 
[ c_3(t) ( \dot \phi_3(t) + 2a \phi_3(t) ) + 
c_1(t) ( \dot \phi_1(t) - a \phi_1(t) ) \nonumber \\ 
&&+ c_2(t) ( \dot \phi_2(t) - a \phi_2(t) )+ 
\pi R^2 \delta  c(t) ( \dot \rho (t) + 2a \rho (t) )] 
- D_0 \int_{- \infty}^{\infty} dt [ {4 \over L^2} ( c_1^2(t) + 
c_2^2(t) \nonumber \\ 
&& + c_3^2(t)) + 2 \pi^2 R^2 ({ R \over L})^6 \delta c^2(t) ] \} 
\ , \ 
\label{af18} 
\eea 
where 
\bea 
&& c_\alpha (t) = {1 \over 2} \epsilon_{\alpha \beta \gamma} 
\int_{\vec x \in \Omega} d^3 \vec x  \hat v_\beta^{>} (\vec x,t) x_\gamma 
\ , \ {\hbox{for $\alpha=1,2$}} 
\ , \ \nonumber \\ 
&& c_3 (t) = \pi R^2 \delta c(t) + {1 \over 2} \int_{\vec x \in \Omega} d^3 \vec x [ \hat v_1^{>} (\vec x,t) x_2 
- \hat v_2^{>} (\vec x,t) x_1 ]\ , \ 
\nonumber \\ 
&& \rho (t) = \omega_3 (t) - \phi_3(t) \ . \ \label{af19} 
\eea
A simplifying prescription has been used to get (\ref{af18}).
The exponential factor $\exp (at)$ has been removed from the expression for $\hat v^{<}_\alpha (\vec x,t)$ and the time integrals have been defined for
$ - \infty < t < \infty $. The point in doing so is that we get gaussian integrals over $\delta c(t)$ and $c_\alpha (t)$, which may be exactly computed. The only consequence of this approximation is just a slight and unimportant deviation for
the values of coupling constants. Taking into account the boundary conditions $\rho ( \pm \infty) = \phi_\alpha ( \pm \infty ) = 0$ in the resulting path-integral, the time variable is then restricted to $-1/a \leq t \leq 0$, where the saddle-point method is assumed to work (this follows naturally from (\ref{msr32}) and (\ref{msr36}), which show that $\hat v^{(0)}_\alpha (\vec x, t)$ and $b(t)$ have lifetimes of the order of $1/a$ and $1/(2a)$, respectively). We will have, therefore, 
\bea 
&Z&( \lambda ) \sim \int_{(D_0 \lambda)^{1/2}}^\infty 
da  \int D[ \rho (t)] \prod_{\alpha=1}^3 
D[ \phi_\alpha (t)] 
\exp  \{ -{{ \pi^2 D_0 R^4} \over {2 L^2}} \cdot {{\lambda^2} \over a} 
- {L^2 \over {4 D_0}} \int_{-1/a}^{0} dt 
[  \dot \phi_3^2 (t) \nonumber \\ 
&+& 4a^2 \phi_3^2 (t) 
+ \dot \phi_1^2 (t) + a^2 \phi_1^2 (t) 
+ \dot \phi_2^2 (t) + a^2 \phi_2^2 (t) 
+ 2 ({L \over R})^4 ( \dot \rho^2 (t) + 
4a^2 \rho^2 (t) )] \} 
\ , \ \label{af20} 
\eea 
an expression which involves a set of uncoupled one-dimensional harmonic oscillators with coordinates $\phi_1$, $\phi_2$, $\phi_3$ and $\rho$.
Observe that $\omega_1(t)$ and $\omega_2(t)$ do not appear in (\ref{af20}).
This means that at length scales of the order of $R$, velocity fluctuations
are essentially axisymmetric. As smaller length scales (of the order of
$R_0$) are considered in the action, vorticity fluctuations
in all directions of space become important. We may write (\ref{af20}) as
\bea 
&Z&( \lambda ) \sim \int_{(D_0 \lambda)^{1/2}}^\infty 
da  \int d \bar \rho d \rho  \prod_{\alpha=1}^3 
d \bar \phi_\alpha d \phi_\alpha 
\exp ( -{{ \pi^2 D_0 R^4} \over {2 L^2}} \cdot {{\lambda^2} \over a} ) 
G( \{ \bar \phi_1| \phi_1 \} ; {1 \over a}, a, { L^2 \over {2 D_0}} ) 
\nonumber \\ 
& \times & 
G( \{ \bar \phi_2 | \phi_2 \} ; {1 \over a}, a, { L^2 \over {2 D_0}} ) 
G( \{ \bar \phi_3 | \phi_3 \} ; {1 \over a}, 2a, { L^2 \over {2 D_0}} ) 
G( \{ \bar \rho | \rho  \} ; {1 \over a}, 2a, { L^2 \over {D_0}} 
({L \over R})^4 ) \ , \ 
\label{af21} 
\eea 
where 
\be 
G( \{ x_2| x_1 \} ; T, \omega , m ) \equiv 
\left ( { {m \omega} \over {2 \pi \sinh ( \omega T)}} \right )^{1/ 2} 
\exp \{ -{ {m \omega} \over {2 \sinh ( \omega T)}} 
[ (x_2^2+x_1^2) \cosh ( \omega T) - 2 x_1 x_2  ] \} 
\label{af22} 
\ee 
is the euclidean propagator \cite{feynman} for a particle of mass $m$ moving, in a time interval $T$, under the
harmonic potential ${ 1 \over 2} m \omega^2 x^2$.
The initial and final coordinates are $x_1$ and $x_2$, respectively. We obtain from (\ref{af21}) and (\ref{af22}) the asymptotic result 
\be 
Z( \lambda ) \sim \int_{(D_0 \lambda)^{1/2}}^\infty  da  {1 \over a^2}
\exp ( -{{ \pi^2 D_0 R^4} \over {2 L^2}} \cdot {{\lambda^2} \over a} ) \sim {1 \over \lambda^2} \ . \
\label{af23} 
\ee

A simple way to understand the regularization of the divergent expression (\ref{msr38}) for $Z(\lambda)$ is that the
additional terms in the path integral summation, associated to fluctuations, are complex quantities, which  
produce an increasing number of canceling factors as $a \rightarrow \infty$.

\section{Subleading corrections}
The asymptotic result (\ref{af23}) does not give us any dimensional parameter which could characterize in a 
more detailed way the circulation PDF, providing further motivation for a comparison with the experiment. 
We will investigate this problem here, through the analysis of subleading corrections for $Z(\lambda)$.

Recalling what has been done, we observe that to derive expression 
(\ref{af23}) the path-integral for $Z( \lambda)$ has been written in a form which depends on an ordinary integral over $a$. The integrand is obtained from the saddle-point method, yielding a consistent result only in the time interval $-1/a \leq t \leq 0$. In this way, fluctuations of the velocity field were completely neglected for $t \leq -1/a$ (for $t \geq 0$ they do not contribute to $Z( \lambda)$ due to causality). An improved form for (\ref{af21}) may be found, thus, through the substitution 
\be 
G(  \{ x_2| x_1 \} ; T, \omega , m ) \rightarrow 
{\rm P}(x_1) G( \{ x_2| x_1 \} ; T, \omega , m ) \ , \ \label{sc1} 
\ee 
where ${\rm P}( x_1)$ is the probability density to have $x=x_1$ at time $t_1 = -1/a$. In other words, the effects of velocity fluctuations for $t \leq -1/a$ are simply encoded in the PDFs for $\rho$ and $\phi_\alpha$. It is important to note that these random variables are related to the circulation at different length scales. We may write, in fact, 
\bea 
&\Gamma_R& \equiv 2 \omega_3(t) \pi R^2 = 2 (\rho(t) + \phi_3(t)) \pi R^2 \ , \ \nonumber \\ 
&\Gamma_{R_0}^{(\alpha)}& \equiv 2 \phi_\alpha (t) \pi R_0^2 
\ . \ \label{sc2} 
\eea 
Above, $\Gamma_R$ is the circulation evaluated for a circular loop of radius $R$ in the $xy$ plane, while $\Gamma^{(\alpha)}_{R_0}$ refers in an analogous way to a loop of radius $R_0$ in a plane perpendicular to the unit vector $\hat x_\alpha$. These loops are centered at the origin of the coordinate system. From (\ref{af22}) we see that as $a \rightarrow \infty$ only small fluctuations of $\phi_\alpha$ and $\omega_3$ become important. These fluctuations are associated to the core of the circulation PDF, which is modeled by a gaussian distribution, 
\be 
{\rm P}(\Gamma_r) \sim \exp( - {{\Gamma^2_r} \over {\Delta (r)^2} } ) \ , \
\label{sc3} 
\ee
where ``$r$" gives the length scale. This form of the circulation PDF for small $\Gamma_r$ is a phenomenological ingredient in our analysis, well supported by numerical and real experiments \cite{cao,sreen}. Using (\ref{sc1}-\ref{sc3}) we rewrite (\ref{af21}) as
\bea 
&Z&( \lambda ) \sim \int_{(D_0 \lambda)^{1/2}}^\infty 
da  \int d \bar \rho d \rho  \prod_{\alpha=1}^3 
d \bar \phi_\alpha d \phi_\alpha 
\exp ( -{{ \pi^2 D_0 R^4} \over {2 L^2}} \cdot {{\lambda^2} \over a} ) 
(1 - {{4 \pi^2 R^4} \over \Delta (R)^2} \omega_3^2 - {{4 \pi^2 R_0^4} \over \Delta (R_0)^2} [ \phi_1^2 \nonumber \\ 
&+& \phi_2^2 +\phi_3^2 ]) 
G( \{ \bar \phi_1| \phi_1  \} ; {1 \over a}, a, { L^2 \over {2 D_0}} ) 
G( \{ \bar \phi_2 | \phi_2 \} ; {1 \over a}, a, { L^2 \over {2 D_0}} ) 
G( \{ \bar \phi_3 | \phi_3 \} ; {1 \over a}, 2a, { L^2 \over {2 D_0}} ) 
\nonumber \\ 
& \times & 
G(  \{ \bar \rho | \rho  \} ; {1 \over a}, 2a, { L^2 \over {D_0}} 
({L \over R})^4 ) \ . \ 
\label{sc4} 
\eea
In order to compute (\ref{sc4}), a very convenient simplification of (\ref{af22}) follows from
\bea
&&x^+ \equiv x_1 e^{{{\omega T} \over 2}} - x_2  e^{-{{\omega T} \over 2}}
\ , \ \nonumber \\
&&x^- \equiv x_1  e^{-{{\omega T} \over 2}} - x_2  e^{{{\omega T} \over 2}}
\ , \ \label{sc5}
\eea
which allows us to write
\be
G( \{ x_2| x_1 \} ; T, \omega , m ) \equiv 
\left ( { {m \omega} \over {2 \pi \sinh ( \omega T)}} \right )^{1/ 2} 
\exp \{ -{ {m \omega} \over {2 \sinh ( \omega T)}} 
[ {1 \over 2} (x^+)^2 + {1 \over 2} (x^-)^2  ] \}  \ . \
\label{sc6}
\ee
It is also necessary to define $\omega_3$ and $\phi_\alpha$ in terms of $\rho^+$, $\rho^-$, $\phi_\alpha^+$ and $\phi_\alpha^-$. We have
\bea
&& \omega_3 = \rho + \phi_3 = { 1 \over { 2 \sinh(2)}} [ e^2 ( \rho^+
+ \phi_3^+ ) - e^{-2} ( \rho^- + \phi_3^-) ] \ , \ \nonumber \\
&& \phi_3 = { 1 \over { 2 \sinh(2)}} [ e^2 \phi_3^+ - e^{-2} \phi_3^- ]
\ , \ \nonumber \\
&& \phi_{1,2} = { 1 \over { 2 \sinh(1)}} [ e \phi_{1,2}^+ - e^{-1} \phi_{1,2}^- ]
\ . \ \label{sc7}
\eea
Substituting (\ref{sc6}) and (\ref{sc7}) into (\ref{sc4}), the gaussian integrals may be readily evaluated, giving
\be
Z( \lambda) \sim { 1 \over \lambda^2} (1 -  {\beta^2  \over \lambda^2}) \ , \ \label{sc8}
\ee
where
\be
\beta \simeq ( 16 \sinh (2) )^{ {1 \over 2} } \Delta^{-1}
\simeq 7.6 \Delta^{-1} \ . \ \label{sc9}
\ee
In the computation of (\ref{sc8}) we have assumed that
\be
{ {\Delta (R_0) R^2 } \over { \Delta (R) R_0^2 }} \gg 1 \ , \ \label{sc10}
\ee
which is clearly verified in practice \cite{cao}.

We may interpret (\ref{sc8}) as the asymptotic approximation to the lorentzian $Z( \lambda) \sim ( \lambda^2 + \beta^2)^{-1}$, which leads, on its turn, to the stretched exponential ${\rm P}( \Gamma) \sim \exp ( - \beta | \Gamma |)$. The tail decaying parameter $\beta$ is inversely proportional, therefore, to the width of the PDF's core, $2 \Delta$. This agrees with Migdal's conjecture that ${\rm P}( \Gamma)$ is a function of the scaling variable $ \Gamma / A^{(2k -1) /2k}$, as discussed in the introduction. We would find (\ref{sc9}) once again if we had considered other axisymmetric contours, as two concentric loops of radius $R_1$ and $R_2$, for instance. The PDF's dependence on the minimal area has to be completely contained in $\Delta$, showing that universal features of the circulation PDF are related essentially to the form of its core. The manifestation of universality not only at the tails of PDFs seems to be in fact a property shared by other turbulent systems, as discussed recently in the problem of a passive scalar advected by a random velocity field in one dimension \cite{balko}.

A physical picture that may explain in more concrete terms the core-tail relationship for the circulation statistics, the result of the above computations, is in order. We may imagine that the large scale forces
generate smooth configurations with small vorticity which are then fragmented in the cascade process up to
the inertial range scales.  These are the ``soft" vortices that contribute to the core of the circulation PDF. With some probability, however, these vortices will be found in regions of the fluid characterized by high stretching. Their vorticity will be, thus, strongly enhanced, producing the intermittent configurations, described by the PDF tails. Since longitudinal velocity differences responsible for stretching  do not fluctuate so quickly as the transverse ones related to circulation, the correlations of the soft vortices are transposed to a different range of vorticity. This is the meaning of $\beta \sim \Delta^{-1}$, which implies that the same anomalous exponents determine the tails and the core of the circulation PDF.

It is clear, from the results just obtained, that our task, within the reach of the saddle-point method, is at best 
to establish predictions suitable to experimental test, even if we lack a precise knowledge of $\Delta(R)$, to 
which further and complementary investigations have to be directed. One might suppose that $\Delta(R)$ 
could be derived, at the onset of turbulence, from the viscous limit of the Navier-Stokes equations, in such 
way that the circulation PDF would keep the form of its core, while developing slowly decaying tails. In the 
viscous case, the circulation PDF is indeed gaussian, but $\Delta(R) \sim R^2$ (see appendix C), which is in 
strong disagreement with observations. Thus, we do not expect smooth configurations of the velocity field to 
play any role in determining the core of the circulation PDF, even in situations close to critical Reynolds 
numbers.        

\section{Parity breaking effects} 
Let us study now possible asymmetries between the left and right tails of the circulation PDF, caused by parity breaking external conditions. We will investigate here two simple models (which will be denoted henceforth by A and B, respectively): a fluid in rotation with constant angular velocity $\vec \omega = \omega \hat z$ and a fluid stirred by the force $\tilde f_\alpha (\vec x ,t) = f_\alpha (\vec x,t) + \bar f_\alpha (\vec x)$, where only $f_\alpha (\vec x,t)$ is random, being defined by (\ref{msr2}). The static component $\bar f_\alpha (\vec x)$ is the one responsible for parity breaking effects. In these models we will assume that the core of the circulation PDF is given by a shifted gaussian distribution,
\be
{\rm P}(\Gamma) \sim \exp( - {{ ( \Gamma - \Gamma_0)^2} \over {\Delta^2} } ) 
\ , \ \label{pbe1}
\ee
with $\Gamma_0 \ll \Delta$, and $\Delta$ being the same as in the situation where parity breaking conditions are removed
( $\omega = \bar f_\alpha (\vec x) =0$). To simplify the notation, we took out the scale dependence of $\Gamma$, $\Gamma_0$ and $\Delta$ in (\ref{pbe1}).

\leftline{{\underline {\bf Model A}}}

A turbulent rotating fluid, with angular velocity
$\vec \omega = \omega \hat z$ is described by a slightly different version of the Navier-Stokes equations (\ref{msr1}), which takes into account the presence of non-inertial effects:
\be
\partial_t  v_\alpha + v_\beta \partial_\beta v_\alpha -2 \omega \epsilon_{3 \alpha \gamma} v_\gamma - \omega^2 x^\perp_\alpha = 
-\partial_\alpha P + \nu \partial^2 v_\alpha +f_\alpha 
\ . \ \label{pbe2}
\ee
The centrifugal force $\omega^2 x^\perp_\alpha$ may be absorbed by the pressure term. Following all the steps carried in sec. II, equation (\ref{msr34}) becomes now
\be
\dot b + 2ab - 2a \omega = -2 \pi  D_0 \lambda
\left ( {R \over L} \right )^2 e^{2at} \theta (-t) \ , \ \label{pbe3}
\ee
which is solved by
\be
b(t)= \omega -{{ \pi D_0 \lambda} \over {2a}} 
\left ( {R \over L} \right )^2 e^{-2a |t|} \ , \ \label{pbe4}
\ee
while equation (\ref{msr23}) still yields the same solution for $\hat v_\alpha (\vec x,t)$, given by (\ref{msr32}) (this is also true for model B; the distinction between the models is due only to different solutions for
$b(t)$). Using (\ref{pbe1}) and (\ref{pbe4}), we obtain the corrected form of (\ref{sc4}), which gives, after computations are done,
\be
Z( \lambda) \sim \exp(-i \lambda \omega ) { 1 \over \lambda^2}
(1 -  \exp( - 2{ {\Gamma_0^2} \over {\Delta^2}}){\beta^2 \over \lambda^2}) \ , \ \label{pbe5}
\ee
We find immediately from (\ref{pbe5}) the shift $\Gamma \rightarrow \Gamma + \omega$ in the circulation PDF, as expected on physical grounds. Another consequence of (\ref{pbe5}) is that the tail decaying parameter $\beta$ gets multiplied by a factor which is related to the shift $\Gamma_0$ at the core of the circulation PDF. As $\Gamma_0$ increases, the PDF tails become broader, apart from the overall shift by $\omega$.

\leftline{{\underline {\bf Model B}}}

Expanding the static part of $\tilde f_\alpha (\vec x,t)$ in a power series around $\vec x = 0$, we will have, up to first order,
\be
\bar f_\alpha (\vec x) = \bar f_\alpha (0) +
\partial_{[ \beta} f_{\alpha ]} x_\beta + \partial_{ \{ \beta}
f_{ \alpha \} } x_\beta \ , \ \label{pbe6}
\ee
where
\bea
&&\partial_{[ \beta} f_{\alpha ]} = \left. { 1 \over 2} ( \partial_\beta f_\alpha - \partial_\alpha f_\beta ) \right |_{\vec x =0} \ , \ \nonumber \\
&&\partial_{ \{ \beta} f_{ \alpha \} } = \left. { 1 \over 2} ( \partial_\beta f_\alpha + \partial_\alpha f_\beta ) \right |_{ \vec x=0} \ . \ \label{pbe7}
\eea
The above expansion is physically associated to parity breaking mechanisms defined in the integral scales. 
As a conjecture, we expect that the induced modification on the instanton solutions will lead to a model 
independent description of parity breaking effects at the PDF tails.

Let us consider here the case where $\partial_{[ \beta} f_{\alpha ]} \equiv \epsilon_{3 \alpha \beta} f_0$, 
to get equations which are still invariant under rotations around the $z$ axis. The strength of parity 
symmetry breaking is given by the external parameter $f_0$. The first and third terms in the RHS of 
(\ref{pbe6}) are absorbed by the pressure in the Navier-Stokes equations. Similarly to the analysis of 
model A, we write the equation for $b(t)$,
\be
\dot b + 2ab = -2 \pi  D_0 \lambda
\left ( {R \over L} \right )^2 e^{2at} \theta (-t) + f_0
\ , \ \label{pbe8}
\ee
which solution is
\be
b(t)=  { {f_0} \over {2 a}} - {{ \pi D_0 \lambda} \over {2a}} 
\left ( {R \over L} \right )^2 e^{-2a |t|} \ . \ \label{pbe9}
\ee
From this we obtain, instead of (\ref{msr37}),
\be
\tilde S^{(0)} = {{ \pi^2 D_0 R^4} \over {2 a L^2}} \cdot
\left \{ (\lambda + i\bar \beta)^2 + \bar \beta^2 \right\}
\ , \ \label{pbe10} 
\ee 
where the $\pi / 2$ rotation $\lambda \rightarrow i \lambda$ was taken into account, and we have
\be
\bar \beta = { { f_0 L^2 } \over { \pi D_0 R^2 } } \ . \ \label{pbe11}
\ee
The result (\ref{pbe10}) may be quickly derived if we note that the only implication of (\ref{pbe9}) is the 
shift $\Gamma \rightarrow \Gamma + \pi R^2 f_0 /a$ in the MSR action, leading to $\tilde S^{(0)}
\rightarrow \tilde S^{(0)} + i \lambda \pi r^2 f_0/a$.

Using now (\ref{pbe1}) and (\ref{pbe10}) to correct (\ref{sc4}), we get,
through a direct computation,
\be
Z( \lambda ) \sim { 1 \over { (\lambda + i \bar \beta)^2 + \bar \beta^2}}
- \exp( - 2{ {\Gamma_0^2} \over {\Delta^2}})
{ \beta^2 \over { [ (\lambda + i \bar \beta)^2 + \bar \beta^2 ]^2 } }
\ . \ \label{pbe12}
\ee
From the above expression for $Z( \lambda)$ we find that the right and left tails of the circulation PDF are described by
$P_+ ( \Gamma ) \sim \exp ( - \beta_+ | \Gamma | )$ and $P_- ( \Gamma)
\sim \exp ( - \beta_- | \Gamma | )$, respectively, with
\bea
&&\beta_+ = \bar \beta + [ \exp( - 2{ {\Gamma_0^2} \over {\Delta^2}})
\beta^2 + \bar \beta^2 ]^{1 \over 2} \ , \ \nonumber \\
&&\beta_- = - \bar \beta + [ \exp( - 2{ {\Gamma_0^2} \over {\Delta^2}})
\beta^2 + \bar \beta^2 ]^{1 \over 2} \ . \ \label{pbe13}
\eea
It is interesting to note that the product of the tail decaying parameters is approximately constant:
\be
\beta_+ \cdot \beta_- = \exp( - 2{ {\Gamma_0^2} \over {\Delta^2}})
\beta^2 \simeq \beta^2 \ . \ \label{pbe14}
\ee
There is a compensation effect between the left and right tails, as the parity breaking parameter $f_0$ is varied.

\section{Conclusion} 
The problem of circulation statistics in fully developed turbulence was
investigated through the Martin-Siggia-Rose formalism. An infinite set of
axisymmetric instanton solutions follows from the saddle-point equations,
which are labelled the component $\sigma_{z z}$ of the strain field, a
partially annealed variable. In physical terms, this means that the non-diagonal
components of the strain tensor, related to circulation, are in fact the random
variables which fluctuate against the quasi-static background defined by
$\sigma_{z z}$. The asymptotic behavior of
$Z( \lambda) = < \exp ( i \lambda \Gamma) >$, as well as its subleading
correction, were found, leading to a stretched exponential description of the
tails of the circulation PDF, a result in agreement with observational data.
The core and the tails of the circulation PDF were seen to be intrinsically
related. We estimate the tail decaying parameter $\beta$ to be approximately
equal to $7.6 \Delta^{-1}$, with $2 \Delta$ being the width of the PDF's
core. The numerical value in this estimate is related to the transition at time
$t \sim - 1/a$ between the saddle-point dominated regime and the free turbulent
description of the fluid in the MSR formalism, which corresponds to have
$\lambda = 0$ in (\ref{msr6}). More generically, if the transition occurs at
time $t \sim -g/a$, where $g$ may be regarded as an adjustable phenomenological
parameter, then we will have $\beta \simeq 4 \sinh(2g)^{1 \over 2} \Delta^{-1}$.
The relationship between $\beta$ and $\Delta$ implies that universal features of
the circulation statistics are determined essentially by the PDF's core, which,
however, cannot be approached by means of the instanton technique.

Parity breaking effects were also studied, as the ones which occur in rotating systems or in fluids stirred by parity breaking external forces. Well-defined predictions were derived, which we believe are within the reach of present numerical techniques, like the method of direct numerical simulations.

On the theoretical side, the important problem to be addressed in future investigations is just the study of the core of the circulation PDF. It is likely that some explicit characterization of vorticity filaments will be necessary in order to study matters as anomalous exponents associated to intermittency and the minimal area conjecture.
 
\section{Acknowledgements} 
This work was partially supported by CNPq. One of us (F.I.T.) would like to
thank FAPEMIG for partial support.
 
\appendix 
\section{Time-dependent translations} 
The MSR action $S(\lambda)$, eq. (\ref{msr6}), is invariant under the group $T$ of time-dependent translations between coordinate systems, defined through 
\bea 
&&\vec x \rightarrow \vec x'= \vec x - \int_0^t dt \vec u(t) 
\ , \  
v_\alpha (\vec x,t) \rightarrow v_\alpha' (\vec x,t) = 
v_\alpha (\vec x + \int_0^t dt \vec u(t),t) - u_\alpha(t) \ , \ \nonumber \\ 
&&\hat v_\alpha (\vec x,t) \rightarrow \hat v_\alpha' (\vec x,t) = 
\hat v_\alpha (\vec x + \int_0^t dt \vec u(t),t) \ , \
Q(\vec x,t) \rightarrow Q'(\vec x,t) = Q(\vec x + \int_0^t dt \vec u(t),t)
 \ , \ \nonumber \\
&&P(\vec x,t) \rightarrow P'(\vec x,t) = P(\vec x + \int_0^t dt \vec u(t),t) 
+\dot u_\alpha(t) x_\alpha \ . \ \label{a1} 
\eea  
We observe that $T$-symmetry holds in the MSR formalism whenever functionals of the velocity field are defined at a fixed instant of time, being also invariant under usual galilean tranformations ($ \vec u(t) =$ constant). 
 
Suppose we have a solution of the saddle-point equations with 
$v_\alpha (\vec x=0,t) = \psi_\alpha (t)$. A time-dependent translation may be applied to find another solution $v_\alpha' (\vec x,t)$ with $v_\alpha' (\vec x=0,t)=0$, which yields the same saddle-point action. Our task is just to determine $\vec u(t)$ from 
\be 
v_\alpha(\int_0^t dt \vec u(t),t)=u_\alpha(t) \ . \ \label{a2} 
\ee 
A simple iterative procedure may be devised to find $\vec u(t)$. 
To start, we note that equation (\ref{a2}) gives 
\be 
u_\alpha (0) = \psi_\alpha (0) \ . \ \label{a3} 
\ee 
Taking now the time derivative of (\ref{a2}), we get 
\be 
\dot u_\alpha (t) - u_\beta(t)
\left. \partial_\beta v_\alpha (\vec x+\int_0^t dt \vec u(t),t) 
\right |_{\vec x=0}- \left. \partial_{t_1} 
v_\alpha (\int_0^tdt \vec u(t),t_1) \right |_{t1=t} = 0 \ . \
\label{a4} 
\ee 
At $t=0$, we have, therefore, 
\be 
\dot u_\alpha (0) - u_\beta(0) \left. \partial_\beta v_\alpha (\vec x,0) \right |_{\vec x=0}- \left. \partial_{t} 
v_\alpha (0,t) \right |_{t=0} = 0 \ , \ \label{a5} 
\ee 
that is 
\be 
\dot u_\alpha (0) =
\psi_\beta(0) \left. \partial_\beta v_\alpha (\vec x,0) \right |_{ \vec x=0}+ \dot \psi_\alpha (0)  \ . \ \label{a6} 
\ee 
We may proceed in the same way, considering expressions generated at each level of the iteration, to find time derivatives up to any order and use them to construct the Taylor expansion of $u_\alpha(t)$ around $t=0$. 

\section{Description of the vortex sheet}
We are taking fluctuations of $v_\alpha^{(1)} (\vec x,t)$ to have a discontinuity at the surface $\partial 
\Omega$, which encloses $\Omega$, a volume with typical size $R_0$. Note, in first place, that we may 
write
\be
v_\alpha^{(1)} (\vec x,t) = v_\alpha^{<} (\vec x,t) [1-F(\vec x,t)]
+v_\alpha^{>} (\vec x,t) F(\vec x,t) \ , \
\label{B1}
\ee
where $v^<_\alpha (\vec x,t)$ and $v^>_\alpha  (\vec x,t)$ are given by (\ref{af6}) and (\ref{af15}), 
respectively, and
\be
F(\vec x,t) =
\cases{ 1, & if $\vec x \in \Omega$\cr
        0, & otherwise.\cr}
\label{B2}
\ee
The idea now is to investigate the consequences of the incompressibility constraint, $\partial_\alpha 
v^{(1)}_\alpha (\vec x,t)=0$. This and (\ref{B1}) imply that
\bea
&& \partial_\alpha  v_\alpha^{<} (\vec x,t) =
\partial_\alpha v_\alpha^{>} (\vec x,t) = 0  \ , \ \nonumber \\
&& (v_\alpha^{<} (\vec x,t) -  v_\alpha^{>} (\vec x,t)) n_\alpha = 0  \ . \  \label{B3}
\eea
Above, $ n_\alpha = \hat n \cdot \hat x_\alpha$, where $\hat n$ is the unit normal vector pointing outwards 
the surface $\partial \Omega$.
Writing $ n_\alpha = R_{\alpha \beta} x_\beta / |\vec x|$, where $R_{\alpha \beta}$ is a rotation matrix, we 
get, from  (\ref{af6}), (\ref{af15}) and (\ref{B3}),
\be
x_\gamma R_{\gamma \alpha}^{-1} \left [ (\bar a_\alpha (t) - a_\alpha (t))+
(b_{\alpha \beta} (t) - \epsilon_{\alpha \sigma \beta} \omega_{\sigma} (t))x_\beta \right ]= 0
\  .  \  \label{B4}
\ee
This gives $a_\alpha (t)= \bar a_\alpha (t)$ and $R^{-1}_{\gamma \alpha}
(b_{\alpha \beta}(t) - \epsilon_{\alpha \sigma \beta } \omega_\sigma (t)) = M_{\gamma \beta}$, where 
$M=M(\vec x)$ is an antisymmetric matrix. Since there is in any closed surface $\partial \Omega$ at least 
one point where $R_{\alpha \beta} =  \delta_{\alpha \beta}$, we find that $b_{\alpha \beta} (t)$ is also an 
antisymmetric matrix. Therefore, $R_{\alpha \beta}$ is constant on $\partial \Omega$ up to rotations 
around $\vec x$, yielding $\hat n = \vec x /|\vec x|$. To put it in another way, $\Omega$ is
a sphere of radius $R_0$. A convenient expression for $b_{\alpha \beta}(t)$ is
\be
b_{\alpha \beta} (t) = \phi_\gamma (t)
\epsilon_{\alpha \gamma \beta} \ , \ \label{B5}
\ee
allowing us to define (\ref{af16}).

\section{Circulation PDF in the viscous limit}
To study the viscous limit, we just neglect the convection term in the Navier-Stokes equations. As a result, we get an instructive example where the circulation PDF may be exactly found. The saddle-point equations (\ref{msr9}) and (\ref{msr10}) are now replaced by
\bea
&&i ( \partial_t v_\alpha - \nu \partial^2 v_\alpha )
= \int d^3 \vec x D_{\alpha \beta} ( |\vec x -\vec x'|) \hat v_\beta (\vec x',t) \ , \ \label{vl1} \\
&&i ( \partial_t \hat v_\alpha + \nu \partial^2 \hat v_\alpha )
= \lambda \epsilon_{3 \beta \alpha} { x_\beta \over r_\perp}
\delta (r_\perp - R) \delta (z) \delta(t) \ . \ \label{vl2}
\eea
The incompressibility constraints $\partial_\alpha v_\alpha = \partial_\alpha \hat v_\alpha =0$ have also to be satisfied. Using (\ref{vl1}) and (\ref{vl2}), the saddle-point action in the MSR functional may be written as
\be
S( \lambda ) = - {\lambda \over 2} \oint_c \vec v \cdot d \vec x \ . \ 
\label{vl3}
\ee
All we need to do, therefore, is to find $v_\alpha (\vec x_\perp, z=0,t=0)
\equiv v_\alpha (\vec x_\perp,0)$. Applying
$( \partial_t  + \nu \partial^2 )$ on equation (\ref{vl1}), we will have, integrating by parts and using (\ref{vl2}),
\be
[ \partial_t^2  - \nu^2 (\partial^2)^2] v_\alpha (\vec x,t) = - F_\alpha (\vec x,t)
\ , \ \label{vl4}
\ee
where
\bea
&&F_\alpha (\vec x,t) = - \lambda \int d^3 \vec x' D_{ \alpha \beta}
(|\vec x - \vec x'|) \epsilon_{3 \gamma \beta} { x_\gamma' \over r_\perp'}
\delta (r_\perp' - R) \delta (z') \delta(t) \nonumber \\
&&\simeq { {D_0 \lambda 2 \pi R^2} \over L^2} \epsilon_{3 \beta \alpha}
x_\beta \exp ( - {{\vec x^2} \over L^2}) \ . \ \label{vl5}
\eea
In Fourier space, equation (\ref{vl4}) becomes
\be
( \omega^2 + \nu^2 k^4 ) \tilde v_\alpha (\vec k, \omega) = \tilde F_\alpha (\vec k) \ . \ \label{vl6}
\ee
We obtain, thus,
\bea
&&v_\alpha (\vec x,t) = { 1 \over { (2 \pi)^2}} \int d^3 \vec k d \omega
{ {\tilde F_\alpha (\vec k)} \over { \omega^2 + \nu^2 k^4}}
\exp ( i \vec k \cdot \vec x + i \omega t ) \nonumber \\
&&= { 1 \over {4 \pi \nu}} \int d^3 \vec k { {\tilde F_\alpha (\vec k)}
\over {\vec k^2}} \exp( i \vec k \cdot \vec x - \nu k^2 |t|)
\ . \ \label{vl7}
\eea
Since we are interested to know $v_\alpha (\vec x_\perp,0)$, it follows, from (\ref{vl7}), that
\be
v_\alpha (\vec x_\perp,0) = { 1 \over {4 \pi \nu}} \int d^3 \vec k { {\tilde F_\alpha (\vec k)} \over {\vec k^2}}
\exp( i \vec k_\perp \cdot \vec x_\perp ) \ . \ \label{vl8}
\ee
Taking now (\ref{vl5}), we get
\bea
&&\tilde F_\alpha (\vec k) =
{ { D_0 \lambda R^2} \over {2 \pi L^2}} \int d^3 \vec x
\epsilon_{ 3 \beta \alpha} x_\beta \exp( -i \vec k \cdot \vec x
-{ {\vec x^2} \over L^2}) \nonumber \\
&&=-i \epsilon_{3 \beta \alpha} k_\beta { {D_0 \lambda \pi^{ {1 \over 2}}
R^2} \over 4} \exp( - { {L^2 \vec k^2} \over 4} ) \ . \ \label{vl9}
\eea
Substituting this result in (\ref{vl8}), we will have
\be
v_\alpha (\vec x_\perp,0) = { { \pi D_0 \lambda R^2} \over {6 \nu}}
\epsilon_{3 \beta \alpha} x_\beta \label{vl10} \ . \
\ee
Thus, from (\ref{vl3}) and (\ref{vl10}), the saddle-point action is computed as
\be
S( \lambda ) = - { {\lambda^2 D_0 \pi^2 R^4} \over {6 \nu}} \ . \
\label{vl11}
\ee
Performing now the analytical mapping $\lambda \rightarrow i \lambda$, we find
\be
Z( \lambda) \propto \exp ( - { {\lambda^2 D_0 \pi^2 R^4} \over {6 \nu}})
\ , \ \label{vl12}
\ee
which leads to a gaussian statistics, described by the circulation PDF
\be
{\rm P}( \Gamma ) = {1 \over { \pi^{ {1 \over 2} } \Delta} }
\exp( - { \Gamma^2 \over \Delta^2} ) \ , \ \label{vl13}
\ee
where
\be
\Delta = ( { {2 D_0} \over { 3 \nu} })^{ {1 \over 2} } \pi R^2 \ . \ \label{vl14}
\ee

\begin{figure} 
\caption{} 
The three axisymmetric surfaces of revolution, I, II and III, which bound 
the support of small scale velocity fluctuations determined by
$\tilde S^{(1)}$. As $a \rightarrow \infty$, the surfaces asymptotically
approach the cone given by $z^2=(x^2+y^2)/8$. 
\end{figure}


\begin{references}
\bibitem{comment} The same idea may be applied to more familiar problems, like flow in a pipe, for instance. In this case, it provides an alternative
derivation of the well-known dependence of the turbulent boundary layer
thickness upon the downstream coordinate and rms fluid velocity.
\end{references}
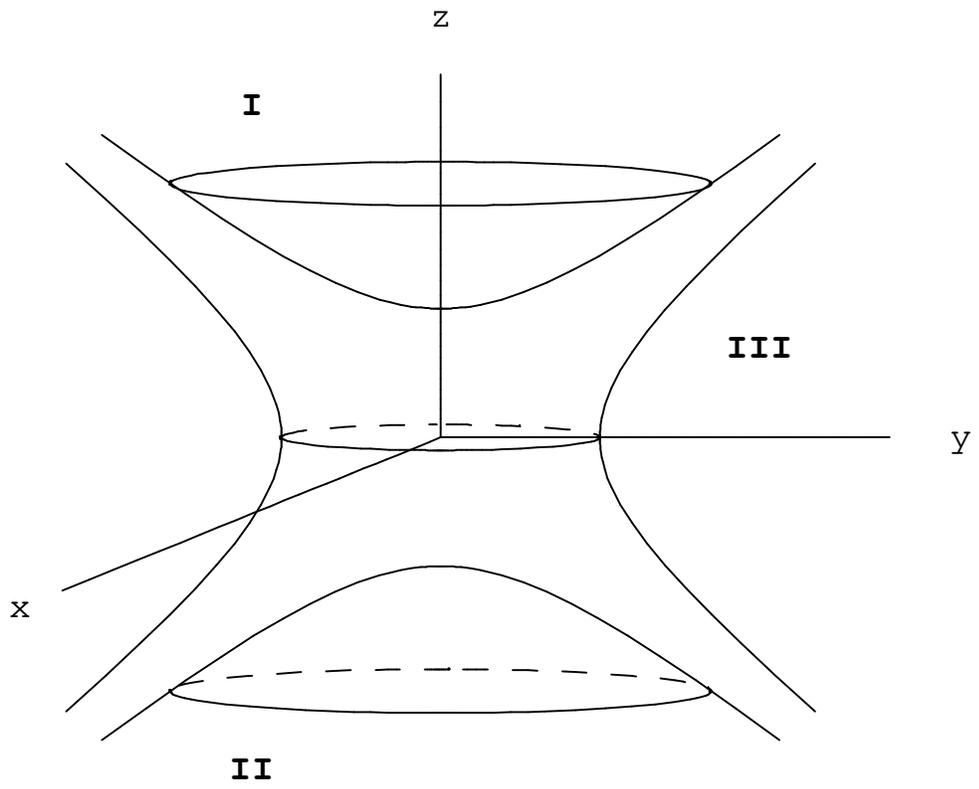
\end{document}